\documentclass[submission,copyright,creativecommons]{eptcs}
\providecommand{\event}{FMAS 2023} 
\usepackage{amssymb}
\usepackage{bm}

\usepackage{graphicx} 
\usepackage{float}
\usepackage[title]{appendix}
\usepackage{amsthm}
\usepackage{mathtools}
\usepackage{stmaryrd}
\usepackage{caption}
\usepackage{subcaption}
\theoremstyle{definition}
\newtheorem{definition}{Definition}[section]
\newcommand{\f}{\mkern-2mu f\mkern-3mu}

\title{Model Checking for Closed-Loop Robot Reactive Planning}
\author{\hspace{-1em}Christopher Chandler
\institute{\hspace{-1em}School of Computing Science\\
\hspace{-1em}University of Glasgow}
\email{christopher.chandler@glasgow.ac.uk}
\and
Bernd Porr
\institute{School of Biomedical  Engineering\\
University of Glasgow}
\email{bernd.porr@glasgow.ac.uk}
\and 
\hspace{2em}Alice Miller
\institute{\hspace{2em}School of Computing Science\\
\hspace{2em}University of Glasgow}
\email{\hspace{2em}alice.miller@glasgow.ac.uk}
\and
\hspace{2em}Giulia Lafratta
\institute{\hspace{2em}School of Engineering\\
\hspace{2em}University of Glasgow}
\email{\hspace{2em}giulia.lafratta@glasgow.ac.uk}
}
\def\titlerunning{Model Checking for Closed-Loop Robot Reactive Planning}
\def\authorrunning{C. Chandler, B. Porr, A. Miller, and G. Lafratta}

\begin{document}
\maketitle

\begin{abstract}
In this paper, we show how model checking can be used to create multi-step plans for a 
differential drive wheeled robot so that it can avoid immediate danger. Using a small, purpose built model checking algorithm \emph{in situ} we generate plans in real-time in a way that reflects the egocentric reactive response of simple biological agents. Our approach is based on chaining temporary control systems which are spawned to eliminate disturbances in the local environment that disrupt an autonomous agent from its preferred action (or \textit{resting state}). The method involves a novel discretization of 2D LiDAR data which is sensitive to bounded stochastic variations in the immediate environment.  We operationalise multi-step planning using invariant checking by forward depth-first search, using a cul-de-sac scenario as a first test case.  Our results demonstrate that model checking can be used to plan efficient trajectories for local obstacle avoidance, improving on the performance of a reactive agent which can only plan one step. We achieve this in near real-time using no pre-computed data. While our method has limitations, we believe our approach shows promise as an avenue for the development of safe, reliable and transparent trajectory planning in the context of autonomous vehicles. 

\end{abstract}

\section{Introduction}\label{sect:intro}
Simple biological systems (or agents) can safely navigate through a previously unseen 
environment by responding in real-time to sensory inputs. On sensing an unexpected input (e.g., an obstacle) the agent responds by performing an action to change its state. This action takes the form of a motor output which results in a change to the environment, which is in turn sensed by the agent and the loop repeats \cite{Braitenberg1986Vehicles:Psychology}. The behaviour is egocentric and reactive---the agent is only concerned with its immediate environment and only deviates from its course (or \emph{resting state}) when necessary. Naturally, complex agents are capable of more sophisticated behaviours, such as the prediction of disturbances and the generation of plans to counteract them. This requires distal sensor information and spatial understanding of the wider environment.  Indeed, there is evidence that in biological systems an
innate ``core" understanding of world physics and causality allow
organisms to organise their behaviours in accordance with predicted outcomes \cite{Spelke2007CoreKnowledge,LeCun2022AOpenReview}.

Model checking \cite{baier_principles_2008} is a widely used technique for automatically verifying reactive systems. It is based on a precise mathematical and unambiguous model of the possible system behaviour. To verify that the model meets specified requirements (usually expressed in temporal logic), all possible executions are checked systematically. If an execution failing the specification is found, the execution path which caused the violation is returned. 
Model checking has previously been successfully used in a variety of different systems. It helped to ensure the safety and reliability of safety-critical systems like flight control \cite{Wang2018},  space-craft controllers \cite{Havelund2001-3} and satellite positioning systems \cite{lu2015}. It has also been  successfully applied to many aspects of software verification, e.g., for industrial systems and operating systems \cite{uppaal-application3, android}.

In autonomous robotic systems, model checking has been used for static and runtime verification \cite{Dennis2016, ferrando2018, cardoso2021, Yang_2022_ros} 
and has been proposed for strategy synthesis \cite{frgihoirmino2020, kwiatkowska2022}. It has been used in many contexts, for example: to generate real-time action plans with formal guarantees for steerable needles
\cite{lehmannneedles2021}); 
industrial robots \cite{weissmann2011}; and for assistive-care robots which  dynamically re-calibrate their path in real-time \cite{hamilton2022}. 

In recent work \cite{pagojus_simulation_2021}, we combined the Spin \cite{holz2011} model checker's ability to identify paths violating temporal properties with sensor information from a 3D Unity simulation of an autonomous vehicle, to plan and perform consecutive overtaking manoeuvres on a traffic-heavy road. The model checker received information from the (simulated) autonomous vehicle, updated its current model, derived a safe path, then communicated the path back to the autonomous vehicle. Although a useful proof-of-concept, the time delay due to model compilation (approx. 3 secs---even though verification of the model itself only took around 20 ms) and the communication between the model checker and game engine was unacceptable. In addition, the requirement to divide the underlying action space into discrete sections made the approach feasible only for less congested environments, such as rural roads. 

In this paper we investigate adapting the approach of \cite{pagojus_simulation_2021} to a real autonomous agent. Our main objective is to demonstrate the effectiveness of two measures which should address the time lag and accuracy problems described above: (i) a simplified model checking algorithm with faster compilation time, and (ii) the use of model checking situated \emph{on board} the autonomous agent. 

In Section \ref{sect:method} we give an overview of our method. Specifically we show how a robot uses multi-step plans derived using model checking to move through a domain, avoiding immediate danger. We present the underlying formal model and describe both how it is used to generate \emph{solution paths} to eliminate disturbances and how the model is updated in real-time. In Sections \ref{sect:implementation} and \ref{sect:results} we describe our implementation and present  results. In Section \ref{sect:discussion} we discuss the implications of our approach---how it compares with previous work, an alternative physics modelling approach, and its current limitations. 

\section{Method}\label{sect:method}

\begin{figure}[t]
\begin{center}
\includegraphics[width=\textwidth]{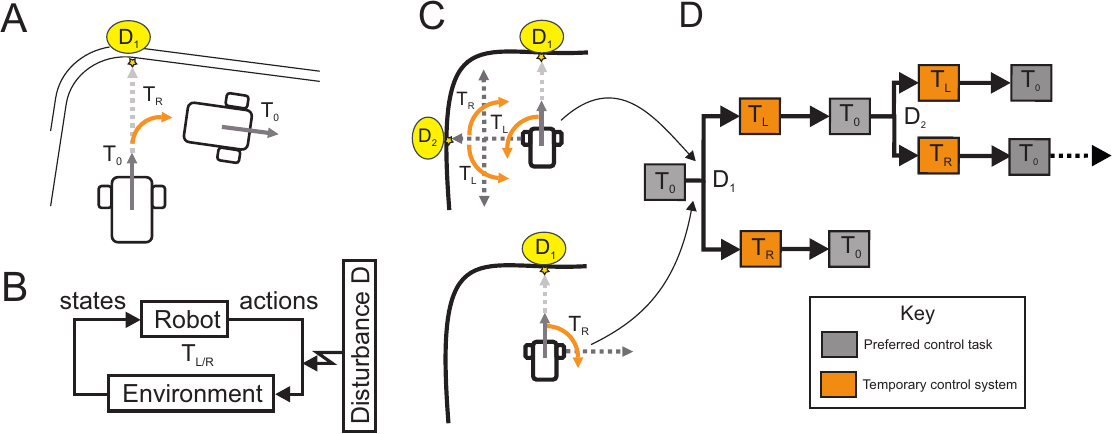}
\caption{Overview of the general concept.  In A the robot is executing its preferred task $T_{0}$ which makes the robot drive in a straight line. The robot approaches a wall and senses a disturbance $D_{1}$ which disrupts the robot agent from its preferred course.  The robot spawns a control task $T_{R}$ (turn right) to counteract $D_{1}$ which can be viewed as the temporary control system shown in B. The task $T_{R}$ only exists until the disturbance is has been counteracted.  However, the disturbance $D_{1}$ could have been eliminated by turning left $T_{L}$. C shows that this would have immediately put the robot in a complex situation which cannot be solved in a stimulus-response fashion by spawning a control task.  In D a reasoning tree is shown for chaining the spawning of temporary control tasks in response to disturbances in the environment.}
\label{fig:concept}
\end{center}
\end{figure}

A typical scenario in which a wheeled robot is driving in an environment avoiding walls and obstacles is shown in Figure\ref{fig:concept} A. The robot agent is initially in a resting state executing a preferred task $T_0$ which makes the robot drive in a straight line. However, from sensory input, the robot can infer that if it continues to follow this path, it will crash into a wall, an unwanted and unexpected event. The robot can predict what is going to happen as it knows its own direction and velocity and can reason about possible courses of action. For example, it can switch to task $T_R$ (turn right) which is executed until the disturbance $D_1$ has been avoided. The task $T_R$ can be viewed as a temporary control system (see Figure \ref{fig:concept}B) with the goal of counteracting the disturbance until it has been avoided, turning states (i.e., sensor data) into actions. Once the disturbance is eliminated, $T_R$ is not needed and the robot returns to task $T_0$. 

While the robot has turned right in Figure 1A, the control goal of counteracting the disturbance $D_1$ could have equally been achieved by turning left, executing task $T_L$.   However, as shown in Figure \ref{fig:concept}C, the robot would have soon encountered a second disturbance $D_2$, a situation which cannot be solved by the temporary control system in Figure \ref{fig:concept}B as it can only reason one step ahead. In Figure \ref{fig:concept}C, if the robot turns left, it can go straight for a while but then has to turn left or right again. Furthermore, if the robot then turns right, it is possible that it might get trapped in a corner, which is undesirable behaviour.  To overcome this limitation, the robot needs some ability to reason about the outcome of chained sequences of tasks for a given number of steps we call the \textit{horizon}.  This reasoning process translates into the tree structure of tasks and sensory inputs shown in Figure \ref{fig:concept}D which forms the foundation of our method.

\begin{figure}[t]
\begin{center}
\includegraphics[width=0.9\textwidth]{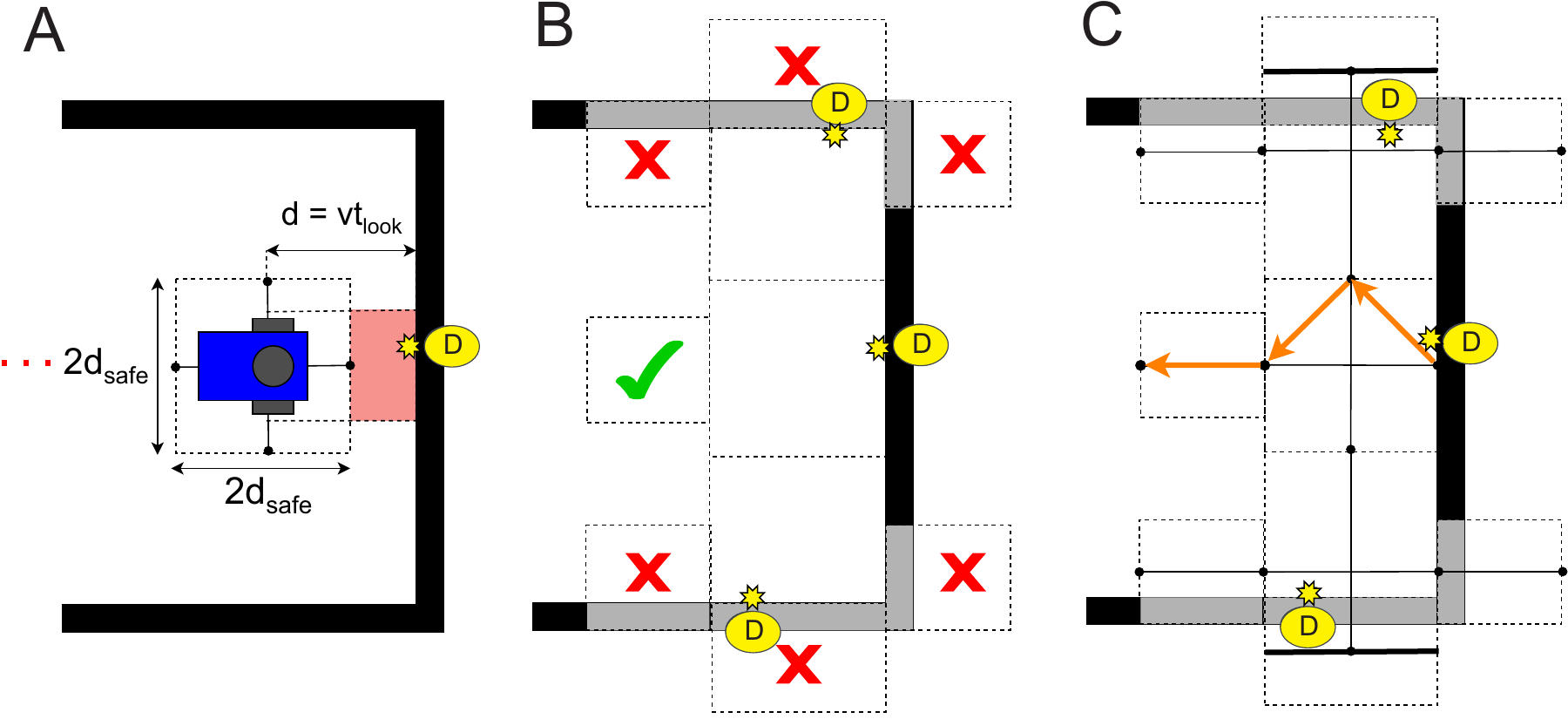}
\caption{Representation of planning sequence for a cul-de-sac.  In A the robot detects a disturbance $D$ (indicated by yellow star), the nearest detectable point associated with the far wall.  A distance $d_{sa{\f}e}$ defines a safe zone so the robot can rotate and acts as a threshold for the robot to start executing plans.  $d = vt_{look}$ is the sensing range derived from the velocity and a lookahead time $t_{look}$. B shows a graphical representation of the model update procedure.  The robot cannot turn right/left then straight nor can it plan an extra step, the only empty set is the one behind.  In C a  path is generated using in situ model checking.  Transitions are extracted and the resulting sequence of tasks (i.e., plan) is executed by control.}
\label{fig:overview}
\end{center}
\end{figure}

Our basic approach is illustrated in Figure \ref{fig:overview}. In Figure \ref{fig:overview}A, a wheeled robot drives towards the far wall of a cul-de-sac in its preferred task $T_0$ and senses a disturbance $D$, operationalised in our case as the nearest sensed point within a distance $d = vt_{look}$ where $v$ is the robot velocity and $t_{look}$ is a set lookahead time. The width of the visual field is determined by the width of the robot plus some tolerance and a check is made for disturbances each iteration of the control loop (approx. every 200 ms). In this scenario, if the robot turns left or right  to avoid the disturbance $D$, the robot will immediately encounter another disturbance.  Furthermore, in either case the robot could get trapped in a corner which is undesirable.

To address this problem, our model is first updated with state information from the environment using a lightweight procedure explained in Section \ref{sect:update}.  In essence, the procedure involves checking whether a proper subset of the robot workspace is empty (i.e., free of obstacles) based on a novel abstraction of the point cloud data.  We utilise symmetry on the axes of a 2D vector space and perform simple filtering to determine whether a given subset is disturbance free in which case the corresponding horizon state is determined \textit{safe}.  Figure \ref{fig:overview}B shows a graphical representation of the procedure outcome, which in this case has updated the model to reflect that turning left twice then returning to $T_0$ is the safe option.  

A valid path for the model is generated utilising a bespoke implementation of model checking (see Figure \ref{fig:overview}C).  We extract transitions from the path to recover the trajectory, which in our case forms a sequence of control tasks from the set $Act = \{T_0, T_L, T_R\}$.  $T_0$ denotes the preferred task of driving in a straight line, and both $T_L$ and $T_R$ are temporary control systems (see Figure \ref{fig:concept}B) which rotate the robot left or right by 90 degrees to eliminate a disturbance.  Real execution is never exact, however error is permissible so long as the control goal of eliminating the disturbance  $D$ is achieved.  The resulting plan is executed when the disturbance is some distance $d < d_{sa{\f}e}$ from the robot.  Once the plan has been executed, the robot returns to $T_0$ until it encounters another unwanted disturbance in the environment.

\subsection{Preliminary model checking}\label{sect:prelim}

\subsubsection{Trajectory specification in LTL}\label{sect:ltl}

We define trajectory specification as the desired sequence of discrete control tasks for an obstacle avoidance scenario, limited in this initial case to static environments.  The planning problem normally consists of two conditions: (i) do not hit any obstacles and (ii) make progress towards a goal \cite{siciliano_springer_2016}.  We address (i) by model checking of a regular safety property. Usually when model checking, counterexamples constitute violation of the property under scrutiny. We however denote any violation a \textit{solution path}.  

In this paper we focus on trajectory specification given as a Linear Temporal Logic (LTL) formula due to its power for discrete sequential planning.  LTL formulae are built from a finite set of atomic propositions $AP$, Boolean connectors such as conjunction $\land$ and negation $\neg$, and two temporal modalities $\varbigcirc$ (``next'') and $\cup$ (``until'') \cite{baier_principles_2008}.  The atomic proposition $a \in AP$ stands for the state label $a$ in a transition system.  For example, in this initial work $AP = \{sa{\f}e, horizon\}$ where $horizon \in AP$ is true in states of the transition system defined as valid planning steps and $sa{\f}e \in AP$ is true in states that can be reached without encountering a disturbance.  The states where $horizon$ is true are known a priori and therefore fixed for our transition system while the states where $sa{\f}e$ is true is decided at runtime. A complete description of our transition system is provided in Section \ref{sect:model}.  

LTL formulae over the set $AP$ of atomic propositions are formed according to the following grammar \cite{baier_principles_2008}: $\varphi ::= \verb|true|\ |\ a\ |\ \varphi_{1} \land \varphi_{2}\ |\ \neg \varphi\ |\ \varbigcirc \varphi\ |\ \varphi_{1} \cup \varphi_{2}$ where a $\varphi$  with an index denotes some arbitrary but distinct formula in LTL. Hence in our case, the atomic propositions $\varphi_1 = sa{\f}e$ and $\varphi_2 = horizon$ are both LTL formulae, so by the grammar the conjunction $\varphi_1 \land \varphi_2$ is also a formula as is its negation $\neg (\varphi_1 \land \varphi_2)$. From this basic grammar, other operators can be derived, such as $\square$ (``always'') and $\lozenge$ (``eventually''), however the derivation is omitted here for brevity (see \cite{baier_principles_2008} for a detailed discussion). 

Our approach utilises a single regular safety property $\square \varphi$ where

\begin{equation}
    \varphi = \neg (sa{\f}e \land horizon)
\end{equation}

\noindent is an invariant expected to hold in each state of the system.  Intuitively, the property $\square \varphi$ says that for any state $s$ in the transition system at least one $a \in AP$ is always false.  As our interest is in \textit{solution paths} not error paths, the invariant $\varphi$ negates the desired outcome, so that what would normally be the set of counterexamples for an infinite run of the system, becomes a set of solutions.  Consequently, the set of solutions paths for our model is the set of paths with a state satisfying the negation of the invariant:

\begin{equation}
    \neg \varphi = sa{\f}e \land horizon
\end{equation}

The set of counterexamples of a regular safety property constitute a language of finite words which can be recognised by a nondeterministic finite automaton (NFA) \cite{baier_principles_2008}.  We therefore construct the NFA $\mathcal{A}_{\square \varphi} = (Q, \Sigma, \delta, Q_{0}, F)$ where $Q$ is a finite set of states, $\Sigma = 2^{AP}$ is a finite alphabet defined as the power set of the $AP$, $\delta : Q \rightarrow 2^{Q}$ is a transition relation, $Q_{0} \subseteq Q$ is a set of initial states, and $F \subseteq Q$ is a set of accept states \cite{baier_principles_2008}.  In fact, for any invariant $\varphi$, the language of all counterexamples (i.e., solutions) can be represented by an NFA with two states.  In our specific case, the NFA $\mathcal{A}_{\square \varphi}$ progresses to the accepting state and terminates if and only if for some state in the transition system the conjunction in (2) is true. 

\subsubsection{Task-driven transition system}\label{sect:model}

A finite transition system is used as a model to describe the behaviour of the robot and provides semantics for trajectory specification in LTL.  The discretized workspace consists of $n$ states $S = \{s_{0}, s_{1},...,s_{n}\}$ and control tasks are interpreted as labelled state transitions to reflect the reasoning tree in Figure \ref{fig:concept}D. 


\begin{definition}[Finite transition system]
A finite transition system $TS$ is a tuple \newline $(S, Act, \rightarrow, I, AP, L)$ where

\begin{itemize}
    \item $S$ is a finite set of states,
    \item $Act$ is a finite set of actions,
    \item $\rightarrow \subseteq S \times Act \times S$ is a transition relation,
    \item $I \subseteq S$ is a set of initial states,
    \item $AP$ is a finite set of atomic propositions, and
    \item $L : S \rightarrow 2^{AP}$ is a labelling function.
\end{itemize}
\end{definition}

\noindent The labelling function $L$ relates a set $L(s) \in 2^{AP}$ of atomic propositions to a state $s$, where $2^{AP}$ denotes the powerset of $AP$ (i.e., the set of all $AP$ subsets including itself and the empty set) \cite{baier_principles_2008}.  Hence the labelling function $L$ assigns truth-values, determining which atomic propositions are satisfied for some state $s$.

The intuitive behaviour of a finite transition system is as follows.  The finite transition system $TS$ starts in some initial state $s_0 \in I$ and evolves according to the transition relation $\rightarrow$.  For convenience, we represent a transition between states with $s \xrightarrow{\alpha} s^\prime$ instead of $(s, \alpha, s^\prime) \in \rightarrow$. If $s$ is the current state, then a transition $s \xrightarrow{\alpha} s^\prime$ originating from $s$ is selected and executed, i.e., the action $\alpha$ associated with the transition is performed, evolving the transition system from state $s$ to $s^\prime$.  In cases where the current state $s$ has more than one outgoing transition, the action $\alpha$ is chosen in a nondeterministic fashion. 

\begin{figure}[t]
\begin{center}
\includegraphics[width=0.9\textwidth]{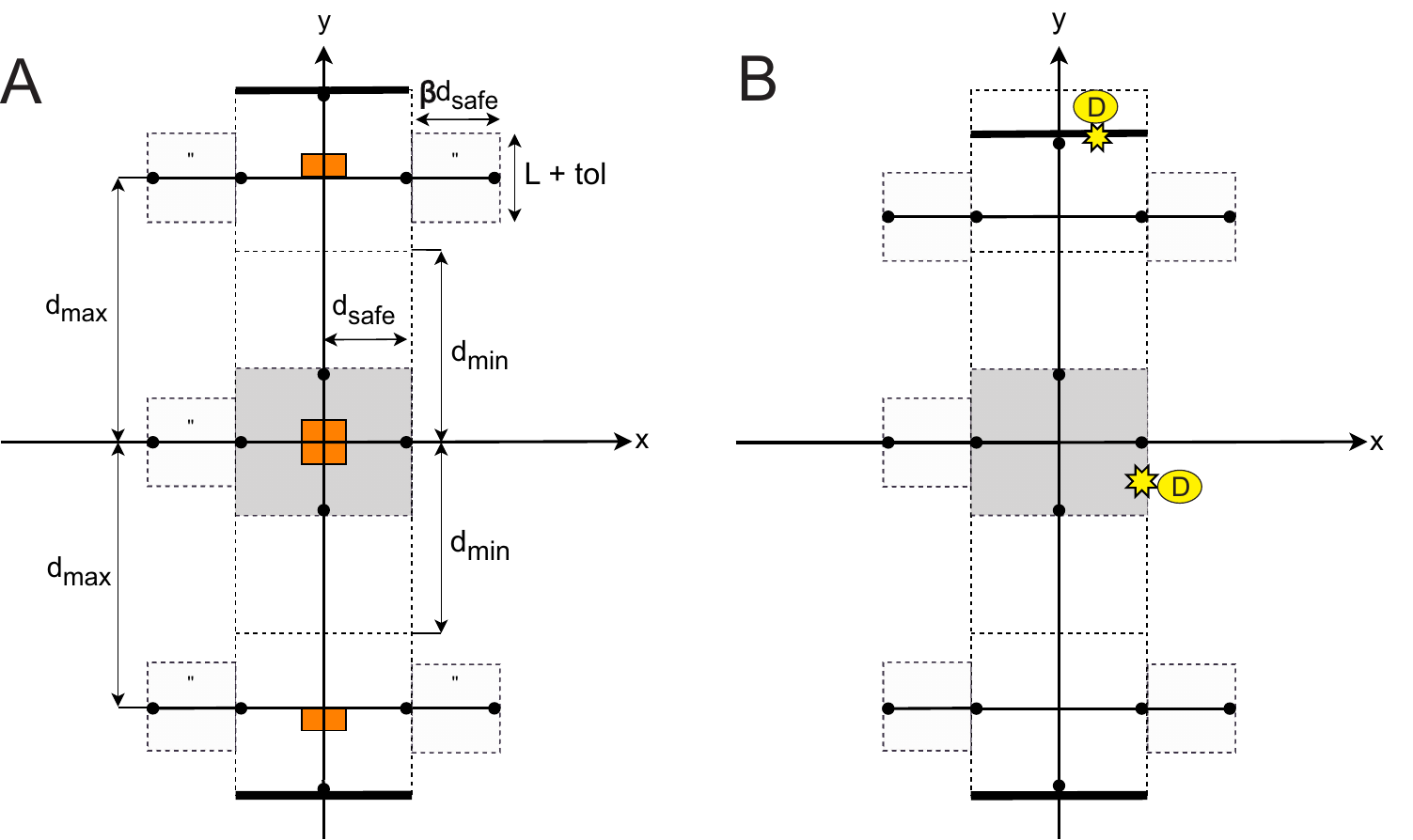}
\caption{Depiction of point cloud abstraction and states. Here A shows the structure of the abstraction where $d_{max}$, $d_{min}$ and $d_{sa{\f}e}$ are tunable parameters constrained by the dimensions of the robot.  $L + tol$ represents the wheelbase of the robot plus some tolerance, which does not need to be as wide as the main driving corridor because the robot can only go straight.  The states (black dots) represent a fixed point distance $d_{sa{\f}e}$ in front of the robot for future robot configurations as it navigates the abstraction, facing towards positive $x$ by convention with the initial state $s_0$ directly in front.  In B an example scenario shows five  states in the positive lateral direction adjusting within the tolerance $d_{max} - d_{min}$ to the location of the nearest disturbance.  Hence lateral states adapt to bounded stochastic variations in the local environment.}
\label{fig:abstract}
\end{center}
\end{figure}

\newpage
Our concrete model of robot behaviour is based upon abstraction of point cloud data from a 2D LiDAR with 360 degrees field of view (see Figure \ref{fig:abstract}A for an illustration).  States $S = \{s_0,s_1,...,s_{14}\}$ represent points on a 2D vector space distance $d_{sa{\f}e}$ in front of possible future robot configurations; the robot faces towards positive $x$ by convention when entering the initial state $s_0$ to generate a plan.   Hence path generation can be seen as the robot reasoning about where it will end up after executing a given sequence of tasks whilst respecting the robot safe zone.  As shown in Figure \ref{fig:abstract}A, the four states on the edge of the safe zone (grey box) represent a fixed point distance $d_{sa{\f}e}$ from the origin, defined to ensure the robot always has adequate clearance for rotational movements when initiating a plan.  In addition, the furthermost state behind the robot on the $x$-axis is also a fixed distance from the origin; this state provides the robot with an option to turn around and go back the way it came if relevant for the scenario.  The $y$-coordinate of states on the lateral extremes of our discretization is variable, however, determined at runtime by the location of disturbances in the positive and negative lateral directions (see Figure \ref{fig:abstract}B for an illustration of the abstraction adjusting to a disturbance in the positive lateral direction).  

As mentioned above, control tasks are interpreted as labelled transitions between states to reflect the desired reasoning tree in Figure \ref{fig:concept}D. Hence we call our model a \textit{task-driven} finite transition system where


\newpage
\begin{itemize}
    \item $S = \{s_0, s_1,...,s_{14}\}$ is a set of states representing a point distance $d_{sa{\f}e}$ in front of the robot for possible future configurations as it navigates the abstraction shown in Figure \ref{fig:abstract},
    \item $Act = \{T_0, T_L, T_R\}$ is the set of discrete control tasks defined in Section \ref{sect:method},
    \item $\rightarrow \subseteq S \times Act \times S$ is a transition relation where $s \rightarrow s^\prime$ is admissible if and only if there exists a control task which can evolve the model from state $s$ to $s^\prime$,
    \item $I = \{s_0\}$ is the initial state distance $d_{sa{\f}e}$ in front of the robot,
    \item $AP = \{sa{\f}e, horizon\}$ is the set of atomic propositions defined in Section \ref{sect:ltl}, and
    \item $L : S \rightarrow 2^{AP}$ is a labelling function such that $L(s)$ determines if property (2) is true at state $s$.
\end{itemize}

\begin{figure}[t]
    \centering
    \includegraphics[width=0.5\linewidth]{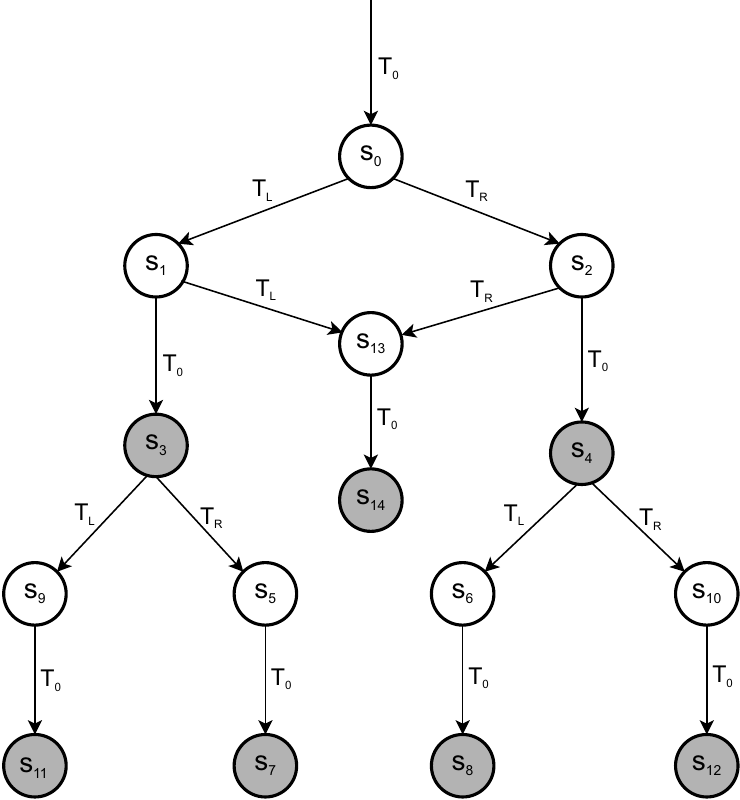}
    \caption{Task-driven finite transition system.  Grey states indicate that $horizon$ is true.  These states are known a priori and fixed for the model. $S_{left} = \{s_{3}, s_{5}, s_{7}, s_{9}, s_{11}\}$ and $S_{right} = \{s_{4}, s_{6}, s_{8}, s_{10}, s_{12}\}$ are determined by the locations of disturbances in the lateral direction.  The remaining states represent fixed points. Transitions reflect the reasoning tree shown in Figure \ref{fig:concept}D for the spawning of control tasks.}
    \label{fig:ts}
\end{figure}

\noindent Figure \ref{fig:ts} shows the structure of our transition system.  In our model, the states where $horizon$ is true are known a priori and therefore fixed, indicated in Figure \ref{fig:ts} by grey states. This is a modelling choice to reflect the final step of any plan, which is to return the robot to its resting state, the preferred task $T_0$.  The transitions to horizon states provide some assurance that the robot has time to replan if another disturbance is encountered soon after plan execution, e.g., due to an error in the real task execution. 

Our model generates a plan of one, two or three steps, as any transition to a safe horizon state (i.e., a state where the property in (2) is true) represents a return to the preferred task $T_0$ and is thus excluded. 
One step plans reflect scenarios where the robot can spawn a single temporary control system $T_{L/R}$ to counteract a disturbance then return immediately to the preferred task $T_{0}$; two step plans are reserved for scenarios where the robot is boxed in and needs to spawn a sequence of two $T_{L/R}$ to about turn and evade the situation; and three step plans add an extra $T_{L/R}$ to counteract disturbances in the lateral direction. 




\subsubsection{Product transition system and NFA}\label{sect:product}

In model checking, we are normally interested in establishing whether $\square \varphi$ is true for all possible runs of a system.  This is equivalent to checking if $Traces_{{\f}in}(TS) \cap \mathcal{A_{\square \varphi}} = \emptyset $ where $Traces_{{\f}in}(TS)$ is the set of finite traces for the transition system \cite{baier_principles_2008}.  To check this, we first construct the product transition system $TS \otimes \mathcal{A_{\square \varphi}}$, then derive an invariant $\varphi$ for the product from the accept states of $\mathcal{A_{\square \varphi}}$ such that  $Traces_{{\f}in}(TS) \cap \mathcal{A_{\square \varphi} = \emptyset}$ if and only if $TS \otimes \mathcal{A_{\square \varphi} \models \square \varphi}$ (i.e., the property $\square \varphi$ is satisfied in the product transition system).  Verification of a regular safety property can therefore be reduced to invariant checking on the product.  However, as we are interested in counterexamples of the property as solution paths, our focus is instead on generating paths in the system for states where the property $\square \varphi$ is false.  

\begin{definition}[Product transition system]
The product transition system $TS \otimes \mathcal{A_{\square \varphi}}$ is a tuple \\ $(S^\prime, Act^\prime, \rightarrow^\prime, I^\prime, AP^\prime, L^\prime)$ where

\begin{itemize}
    \item $S^\prime = S \times Q$. $s^\prime = \langle s, q \rangle \in S^\prime$, $\forall s \in S$ and $\forall q \in Q$, 
    \item $\rightarrow^\prime$ is the smallest relation defined by the rule $\frac{s_{i} \xrightarrow{\alpha} s_{j} \land p \xrightarrow{L(s_{j})} q}{\langle s_{i}, p \rangle \xrightarrow{\alpha}^\prime \langle s_{j}, q \rangle}$,
    \item $I^\prime = \{\langle s_{0}, q \rangle\ |\ s_{0} \in I \land \exists q_{0} \in Q_{0}.\ q_{0} \xrightarrow{L(s_{0})} q\}$.
    \item $AP^\prime = Q$, and
    \item $L^\prime : S \times Q \rightarrow 2^{Q}$ is given by $L^\prime(\langle s, q \rangle) = \{q\}$.
\end{itemize}
\end{definition}

It suffices to perform a reachability analysis on $TS \otimes \mathcal{A_{\square \varphi}}$ to check the invariant $\varphi$.  In this paper, we implement and perform invariant checking by forward depth-first-search (f-DFS) (see Algorithm 4 in \cite{baier_principles_2008} for details).  Our model is updated with state information from sensors at runtime to determine which horizon states are safe and generate a sequence of states from which the associated control tasks can be extracted.  If $\square \varphi$ is false for some state, then the execution path in which the state is reached (normally referred to as a counterexample) is called a \textit{solution path} for the task-driven finite transition system, which in our case reflects a sequence of discrete robot configurations in the workspace.  We then extract the tasks associated with each transition to recover the trajectory for the control layer to execute.

\subsection{Model update procedure}\label{sect:update}

\begin{figure}[t]
    \centering
    \includegraphics[width=0.55\linewidth]{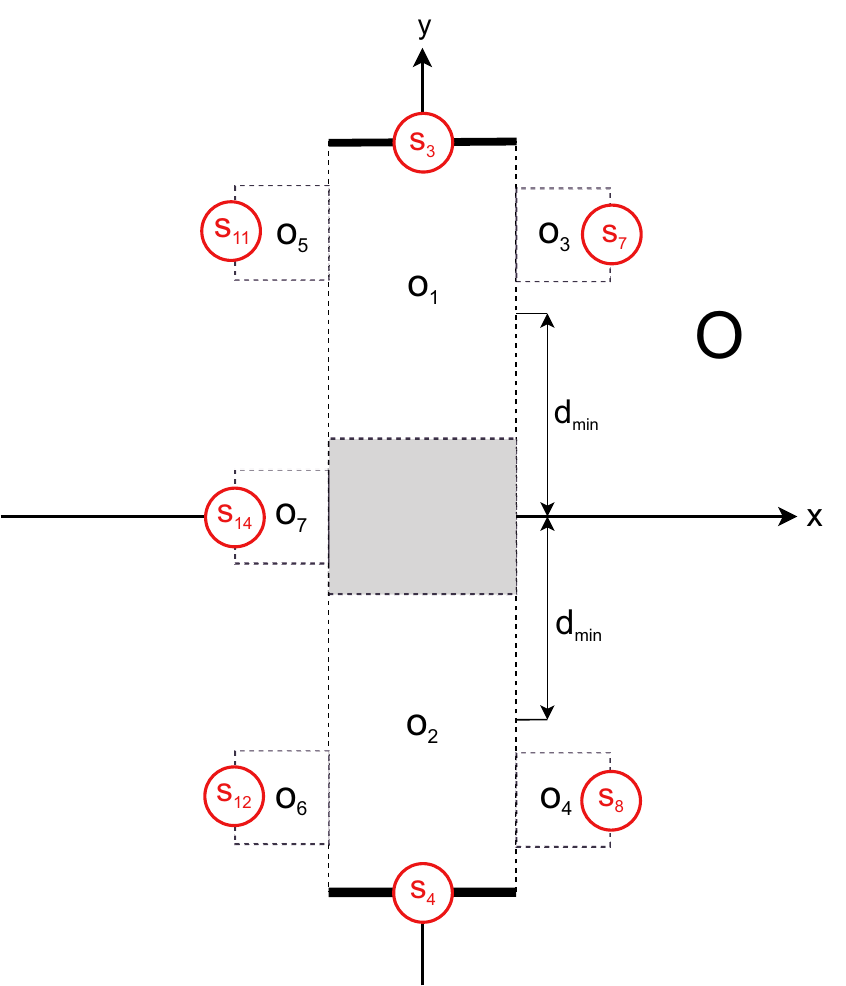}
    \caption{The set of point cloud observations $O$ and the disjoint subsets $o_{1}, o_{2},...,o_{7}$ which form the abstraction.  If some $o_{i} = \emptyset$, the atomic proposition $sa{\f}e$ is true for the horizon state on the edge of the set (indicated with red circles), otherwise the proposition is false. In addition, if the robot cannot travel a lateral distance $d_{min}$ in either direction, then it is assumed  $o_{7} = \emptyset$ to address trap situations, e.g., getting stuck in a corner.  In static environments this is valid as the direction the robot came from should be safe.} 
    \label{fig:sets}
\end{figure}

As mentioned above, states where $horizon$ is true are known a priori (see Figure \ref{fig:ts} for details), so for a solution path to be generated it remains for us to decide which of these states is also safe.  We utilise longitudinal and lateral offsets of the point cloud to simulate respective displacements and take advantage of symmetry on the axes of a 2D vector space to determine if a  subset of our abstraction, specified to represent a practical over-approximation of the task execution workspace, contains no disturbance. If a given subset is empty, the corresponding state is reachable and determined safe, indicated in Figure \ref{fig:sets}.

Our abstraction in Figure \ref{fig:abstract} assumes the robot is at the origin of an underlying 2D vector space used as a model for the point cloud data.  The robot is facing towards positive $x$ by convention with the initial state $s_0$ distance $d_{sa{\f}e}$ directly in front of the robot on the $x$-axis.   In Figure \ref{fig:sets}, subset $o_1$ is a reflection of $o_2$ with the $x$-axis forming a line of symmetry (grey box indicates the robot safe zone, however both sets extend to the $x$-axis). For each of these subsets, the $y$-axis also forms a line of symmetry which splits each subset in half, such that each half is a reflection of the other.  Symmetry on the $x$-axis means we can use a simple inequality to decide whether an observation has a qualifying $y$-coordinate (within the lateral bounds of the abstraction, positive for set $o_1$ and negative for set $o_2$).  Symmetry on the $y$-axis means we can use absolute values to ignore the sign and ensure that any included observation has an $x$-coordinate which does not exceed some maximum distance from the $y$-axis.  For sets $o_1$ and $o_2$, this distance is $d_{sa{\f}e}$, hence the longitudinal dimension of these sets respects the robot safe zone.

When a disturbance $D$ is sensed, the procedure is initiated.  First a longitudinal offset is calculated by subtracting $d_{sa{\f}e}$ from the $x$-coordinate so we can forward simulate our abstraction to its location: 

\begin{equation}\label{eq:offx}
    \triangle_x = D_x - d_{sa{\f}e}
\end{equation}

\noindent However, if $D_x \leq d_{sa{\f}e}$ then $\triangle_x = 0$, as the abstraction is already at the desired displacement from the disturbance in the longitudinal direction, so no forward simulation is necessary for reasoning. 

We then iterate the observations $O$, forward simulate each observation by subtracting $\triangle_x$ from the $x$-coordinate, and use symmetry on the axes of the vector space to sort observations into relevant subsets:

\begin{equation}
   o_1 = \{o \in O\ |\ 0 < o_y < d_{max} + d_{sa{\f}e} \wedge |o_x| \leq d_{sa{\f}e}\}
\end{equation}
\begin{equation}
   o_2 = \{o \in O\ |\ -(d_{max} + d_{sa{\f}e}) < o_y < 0 \wedge |o_x| \leq d_{sa{\f}e}\}
\end{equation}

\noindent where 0 is a constant to distinguish between positive and negative $y$-coordinates, the expression $d_{max} + d_{sa{\f}e}$ represents the absolute distance to lateral extremes of our abstraction, and $|o_x| \leq d_{sa{\f}e}$ ensures the width of the subset respects the robot safe zone. Subsequent reasoning can be seen as the robot predicting what will happen if it executes a one step plan to avoid the disturbance and return to its resting state.

If $o_{1} = \emptyset$ and $o_{2} = \emptyset$, then in either case we can infer that the subset is free of disturbances, meaning that the robot can execute $T_L$ or $T_R$ then return to the preferred task $T_0$ at least for distance $d_{max}$ (a set parameter for the furthermost possible lateral configuration of the robot from the origin of the vector space, as per the point cloud abstraction shown in Figure \ref{fig:abstract}A).  From the robot agent perspective, this means that both states $s_{3}$ and $s_{4}$ can be reached without encountering a disturbance (insofar as it knows) and a path is generated nondeterministically.  If $o_{1} = \emptyset$ and $o_{2} \neq \emptyset$, then the robot can only execute $T_L$ before returning to $T_0$, so a path is generated for $s_{3}$ not $s_{4}$.  If $o_{1} \neq \emptyset$ and $o_{2} = \emptyset$, then the robot can only execute $T_R$ before returning to $T_0$, so a path is generated for $s_{4}$ not $s_{3}$. Where any of these conditions hold, a one step plan is generated without progressing the procedure, as the robot would like to return to its resting state as soon as possible. However, if both $o_1$ and $o_2$ are not empty, we can conclude that the robot will soon encounter another disturbance once it returns to $T_0$, so a one step plan is not possible. 

Next a two step plan is considered, i.e.,  whether the robot is boxed in and should turn 90 degrees twice to go back the way it came, or has enough room in the lateral directions to execute a three step plan. The nearest positive lateral disturbance $D^+ = \min | o_y |$ for $o \in o_1$ and nearest negative lateral disturbance $D^- = \min |o_y|$  for $o \in o_2$ is acquired.  As long as $|D^+_y|$ or $|D^-_y|$ is greater than $d_{min}$, it is concluded that the robot can travel at least distance $d_{min} - d_{safe}$ in the associated lateral direction.  It is therefore considered a safe initial direction and a valid three step plan exists for the scenario. Otherwise, the robot infers that it is boxed in, so the only empty set is $o_{7}$ (immediately behind the robot), leading to the generation of a two step plan in which the robot turns around to go back the way it came (assumed to be safe).  In this case, a two step plan is generated and the procedure terminates, as further reasoning is unnecessary.

However, if at least one direction is determined initially safe for a three step plan, positive and negative lateral offsets are calculated so that we can reason about counteracting any lateral disturbances:

\begin{equation}\label{eq:offxpos}
    \triangle^+_y = D^+_y - d_{sa{\f}e}
\end{equation}
\begin{equation}\label{eq:offyneg}
    \triangle^-_y = D^-_y + d_{sa{\f}e}
\end{equation}

\noindent where $\triangle^+_y$ estimates the maximum lateral displacement of the robot in the positive direction while respecting the robot safe zone, and  $\triangle^-_y$ represents the same for the negative direction.  At this point in the procedure, we have no more use for sets $o_1$ and $o_2$ so they do not participate in any further reasoning.  Instead we build sets $\{o_3, o_5, o_4, o_6\}$ to reason about counteracting any lateral disturbances. Progressing this far means that we have already reasoned about the initial two steps the robot can execute. 

Lateral offsets translate the point cloud data, such that the robot remains at the origin and the axes of the vector space again form lines of symmetry on the relevant subsets.  For example, when $\triangle^+_y$ is applied, the robot is at the origin distance $d_{sa{\f}e}$ from the nearest disturbance on the left of the robot, representing an egocentric perspective of its future location if it first executed the sequence $\langle T_L, T_0 \rangle$ (the orientation of the robot is of course different, however for our purposes this can be ignored).  As a result, states $s_{11}$ and $s_7$, which are by design a fixed lateral distance $d_{sa{\f}e}$ from any sensed lateral disturbances, sit on the $x$-axis, such that $s_{11}$ is directly behind the robot and $s_{7}$ is directly in front.   The states are a midpoint in the lateral dimension for the associated subsets, hence $o_5$ is behind the robot, $o_3$ is in front, and the $x$-axis forms a line of symmetry which splits each subset in half.   This means we can use absolute values to ignore the sign and ensure that any included observation has a $y$-coordinate that does not exceed some maximum distance from the $x$-axis.  Symmetry on the $y$-axis means we can use a simple bounded inequality to decide whether an observation has a qualifying $x$-coordinate (whilst excluding any of subsets $o_1$ and $o_2$).

We therefore iterate $O$, forward simulating each observation subtracting $\triangle_x$ from the $x$-coordinate as before, but adjusting for the lateral displacement by subtracting $\triangle^+_y$    or $\triangle^-_y$ from the $y$-coordinate.  $O$ is iterated once for each direction. In either case, the lateral offset places the relevant three step horizon states on the $x$-axis so we can use symmetry on axes of the vector space for specifying relevant subsets:

\begin{equation}
   o_3, o_4 = \{o \in O\ |\ d_{sa{\f}e} < o_x \leq \beta d_{sa{\f}e} \wedge |o_y| \leq \frac{1}{2}(L + tol)\}
\end{equation}
\begin{equation}
   o_5, o_6 = \{o \in O\ |\ -\beta d_{sa{\f}e} \leq o_x < -d_{sa{\f}e} \wedge |o_y| \leq \frac{1}{2}(L + tol)\}
\end{equation}

\noindent where $d_{sa{\f}e}$ is a constant to distinguish between positive and negative $x$-coordinates (while excluding subsets $o_1$ and $o_2$), $\beta$ is a coefficient for tuning the length of the subsets and $\frac{1}{2}(L + tol)$ defines the width (see Figure \ref{fig:abstract}A for an illustration).  From a cognition perspective, subsequent reasoning can be seen as the robot predicting what will happen if it returns to its resting state after eliminating a second disturbance.

Suppose the positive and negative lateral directions have both been calculated as safe for a three step plan, i.e., both $|D^+_y|$ and $|D^-_y|$ are greater than $d_{min}$.  If subsets $\{o_3, o_5, o_4, o_6\}$ are empty, then states $\{s_7, s_{11}, s_8, s_{12}\}$  are deemed safe and a path is generated for one of the states in a nondeterministic way.  However, if any of the subsets are non-empty, the corresponding state is unsafe and excluded from path generation via model checking. If $|D^+_y| \leq d_{min}$, then states $s_7$ and $s_{11}$ are automatically considered unsafe (i.e., the positive lateral direction is invalid for a three step plan). If $|D^-_y| \leq d_{min}$, then states $s_8$ and $s_{12}$ are automatically considered unsafe (i.e., the negative lateral direction is invalid for a three step plan). As mentioned above, prior to generating a three step plan, if $|D^+_y|$ and $|D^-_y|$ are less than or equal to $d_{min}$, neither direction is considered safe, so a two step plan to turn around and evade the situation is generated.

\section{Implementation}\label{sect:implementation}

\begin{figure}[t]
    \centering
    \includegraphics[width=\linewidth]{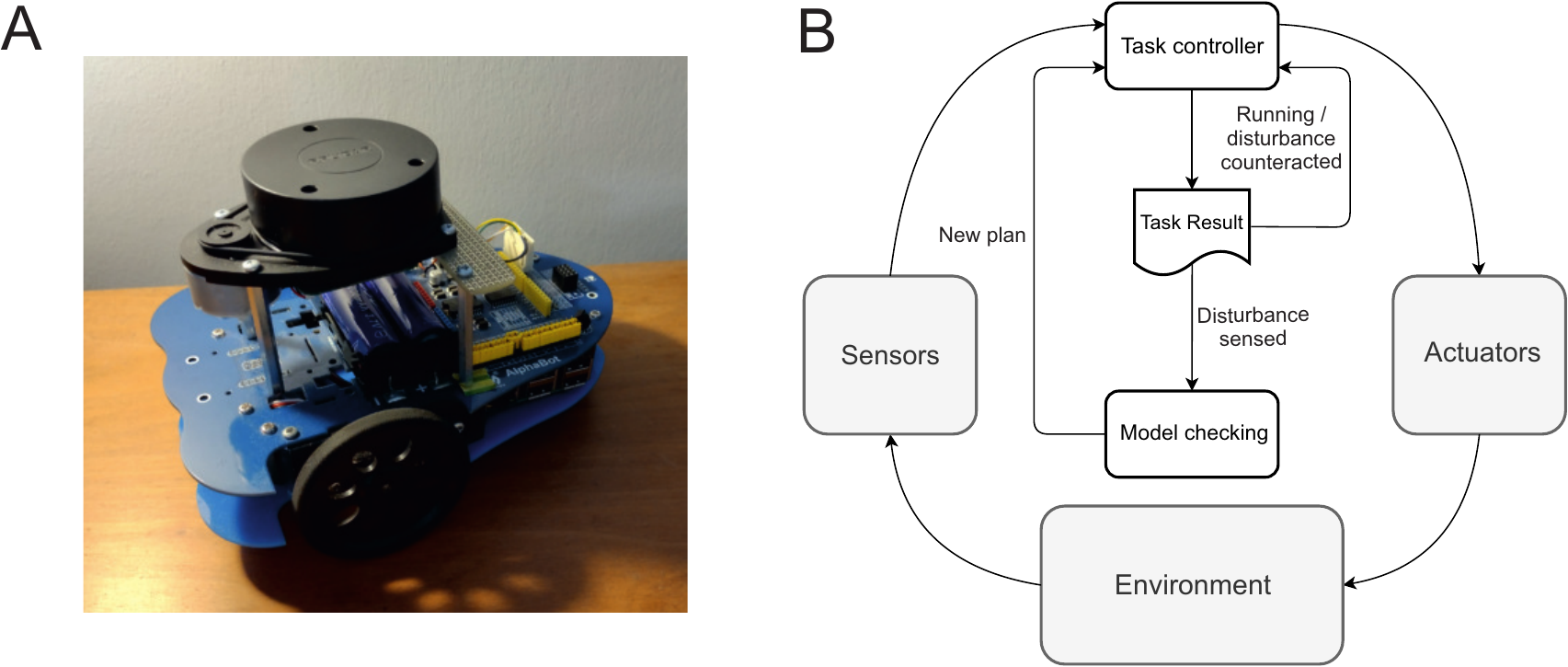}
    \caption{A our robot.  B the agent architecture.}
    \label{fig:arch}
\end{figure}

As a case study, we implemented our method on a differential drive robot shown in Figure \ref{fig:arch}A.  Our robot was adapted from a widely available mobile robot development platform, AlphaBot by Waveshare\footnote{\url{https://www.waveshare.com/alphabot-robot.htm}}.  For sensing the environment, we equipped the robot with a low cost 360 degree 2D laser scanner, RPLiDAR A1M8 by Slamtec\footnote{\url{https://www.slamtec.com/en/LiDAR/A1/}}, and for actuation we used two continuous rotation servos by Parallax\footnote{\url{https://www.parallax.com/product/parallax-continuous-rotation-servo/}}.  The hardware programming interface for the robot was a Raspberry Pi 3 Model B\footnote{\url{https://www.raspberrypi.com/products/raspberry-pi-3-model-b/}} included with the AlphaBot development kit running a Quad Core 1.2GHz Broadcom 64bit CPU with 1GB RAM and wireless LAN.  

Our method was implemented\footnote{\url{https://github.com/possibilia/mc-avoid}}
in C++ using the closed-loop agent architecture shown in Figure \ref{fig:arch}B.  At runtime, LiDAR scans generate a callback from a dedicated thread which passes observations to the agent event handler (approx. every 200 ms).  The agent then initiates the task execution step which sends a control signal to the actuator thread.  If the robot is in the preferred task $T_0$ after the control signal is sent to the servos, a check is made for new disturbances in the environment.  If a temporary control system $T_{L/R}$ is the current task, the progress of counteracting the disturbance is checked.  In either case, a task result is returned to the agent indicating whether the task has been a success or has failed.  When in the preferred task $T_0$, if there is no plan available, a new disturbance initiates the model update procedure described in the previous section and generates a plan using invariant checking by f-DFS.  The plan is executed when the disturbance is distance $d < d_{sa{\f}e}$ from the origin of the point cloud data.  Once the plan has been executed, the robot defaults to the preferred task $T_0$ until a disturbance repeats the process.

\section{Results}\label{sect:results}

Our goal was to develop an egocentric method for chaining temporary control systems in response to disturbances in the environment using in situ model checking.  Specifically, we were interested in improving on the case where an agent can only spawn a single control task in response to disturbances (see Section \ref{sect:method} for details).  In this paper, we restricted our attention to static environments and local tactical planning for avoiding obstacles, focusing on a cul-de-sac scenario as an initial test case for our method.  Our results show an improvement on one step planning yielding efficient trajectories for avoiding a cul-de-sac.  In addition, for both comparisons our model checking procedure was executed in less than 11 ms.

\begin{figure}
    \centering
    \includegraphics[width=0.8\linewidth]{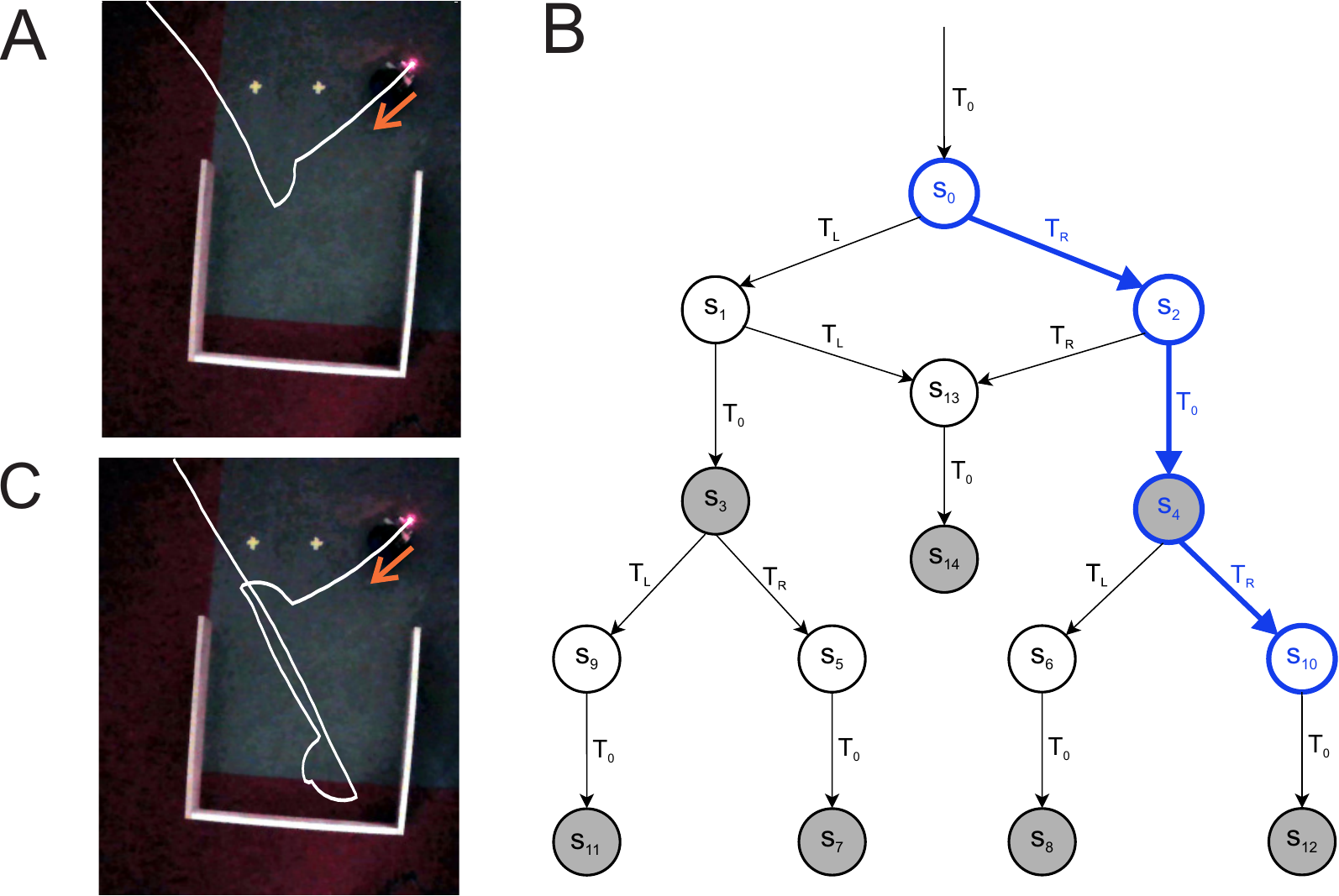}
    \caption{Comparison 1.  In A the robot recognises that it would get boxed in turning left so it makes a three step plan for the right direction (last step cropped).  In B the path and transitions in the model which generated the behaviour are shown. C shows trajectory followed by agent which can only plan one step.}
    \label{fig:comparison1}
\end{figure}

\begin{figure}
    \centering
    \includegraphics[width=0.8\linewidth]{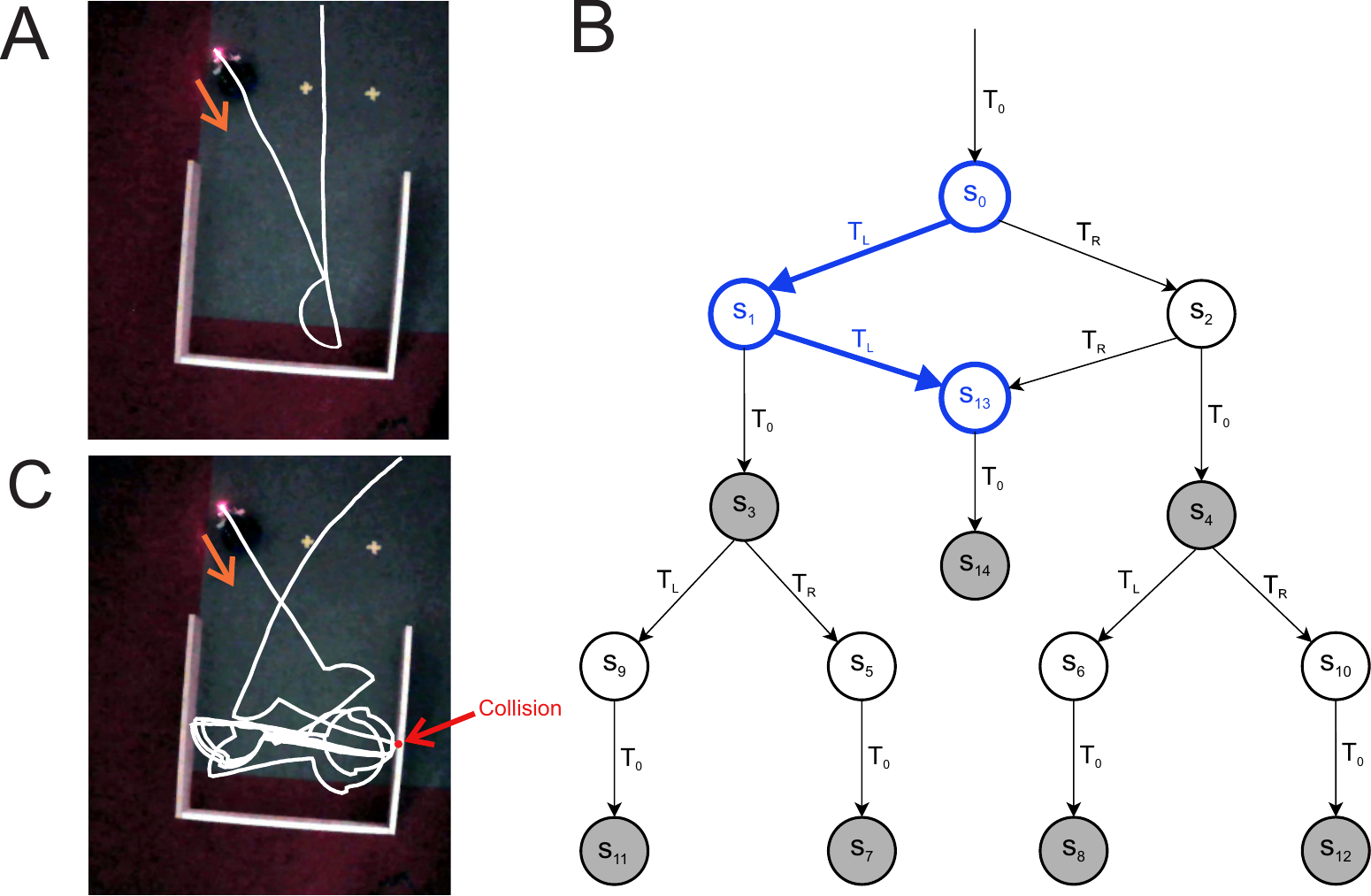}
    \caption{Comparison 2. In A the robot approaches the bottom right corner of the cul-de-sac and infers that it is boxed in so generates the two step plan shown in B to evade the situation.  C shows that the robot gets trapped between two walls for the one step planning case, eventually colliding with a wall.}
    \label{fig:comparison2}
\end{figure}

Trajectories for the the first comparison are shown in Figure \ref{fig:comparison1}.  Using our method, the robot approaches the cul-de-sac in Figure \ref{fig:comparison1}A and upon sensing a disturbance in the environment (i.e., a point on the wall on the left) generates the three step plan $\langle T_R, T_0, T_R \rangle$ in Figure \ref{fig:comparison1}B. Here the model update procedure determines that sets $o_{1}$ and $o_{2}$ (see Figure \ref{fig:sets} above) are non-empty and that it can drive at least $d_{min} = 0.5$m in either direction.  However, as sets $o_{3}$ and $o_{5}$ are also non-empty, it infers that it will meet a disturbance if it turns either left or right after initially going in the left direction.  In this particular case, both sets $o_{4}$ and $o_{6}$ are empty and so the corresponding horizon states $s_{8}$ and $s_{12}$ are determined safe.  A path to one of the states is chosen nondeterministically, in this case $s_{12}$.  Execution time for the combined model update procedure and path generation using in situ model checking was 9.22 ms.  Figure \ref{fig:comparison1}C shows the one step comparison, which in this case manages to navigate out of the cul-de-sac by chance, however not without entering it first, making the trajectory less efficient.

The second comparison is shown in Figure \ref{fig:comparison2}.  In this case, the robot approaches the bottom right corner of the cul-de-sac in Figure \ref{fig:comparison2}A and infers that it is boxed in using our method.  Consequently, it generates the two step plan $\langle T_L, T_L \rangle$ in Figure \ref{fig:comparison2}B then returns to the preferred task $T_0$.  As in the previous comparison, the model update procedure has determined that sets $o_{1}$ and $o_{2}$ are non-empty, however in this case the robot cannot travel at least $d_{min} = 0.5$m in any direction, so a three step plan is invalid; it is therefore assumed that only set $o_{7}$ is empty and the corresponding horizon state $s_{14}$ is safe.  As there is only one safe horizon state in this situation, nondeterminism is resolved in the algorithm so the path is fully determined. Similar to the previous comparison, execution time for the combined update procedure and path generation was 10.87 ms.  In contrast, the one step planning trial causes the robot to get trapped between two walls for a significant period of time before eventually colliding with one of them, as shown in Figure \ref{fig:comparison2}C.  Hence our method generates a more efficient trajectory and can avoid a trap situation.

\section{Discussion}\label{sect:discussion}

In this paper, we have shown that it is possible to use live model checking to plan a safe sequence of discrete control tasks for a planning horizon of more than one step. In previous work \cite{pagojus_simulation_2021}, it was demonstrated that model checking could be used to plan overtaking manoeuvres for an autonomous vehicle (AV) as a proof of concept.  As mentioned in the introduction, however, one of the main issues was compilation time in Spin \cite{holz2011} (approx. 3 secs), making the real-world application of model checking for trajectory planning impractical, even though verification of the model itself only took around 20 ms.  We have overcome the compilation time problem by creating a stripped down model checking algorithm situated on board an autonomous agent. While our results are preliminary, we consider this work a successful first step towards real-time model checking for reliable and safe AV trajectory planning.

As finite state model checking relies on a discrete action space, another limitation in \cite{pagojus_simulation_2021} was division of the underlying continuous system into 21 meter long segments, which is sufficient to represent rural roads or empty motorways but not congested urban environments.  While using a fine-grained discretization would allow for more accurate modelling of the environment and vehicle speed, the state-space explosion problem imposes hard practical limits. If the state-space is too large, the additional computation would make real-time application of model checking infeasible, especially in high speed environments such as autonomous driving. In \cite{li2016} it was argued that even a lag of 100 ms is unacceptable. 

We have therefore introduced a novel discretization which moves away from static grids commonly applied in model checking, using abstraction to represent bounded stochastic variations in the continuous system. This has the benefit of keeping the state-space small but the model sensitive to fine-grained variations in the local environment.  In this initial work, the upshot of our discretization is not obvious, however we believe that for any real-world application of model checking for AV trajectory planning, a discrete representation of the environment with adaptive characteristics will be necessary.

One limitation, however, is that our discretization of the LiDAR data models points on a 2D vector space.  While sufficient for initial research, realistic driving scenarios will require richer information about the geometry of the local environment for reliable decision-making. If the dimensionality of objects in the environment is not known, then sophisticated manoeuvres will be impossible.  For example, overtaking can be broken down into three distinct sub-manoeuvres \cite{dixit_trajectory_2018}: (i) lane change to overtaking lane, (ii) pass leading vehicle(s), and (iii) lane change back to original lane, so for a successful overtake, the AV needs to know the necessary lateral and longitudinal displacements relative to the leading vehicle(s).  Points on a plane have no dimensionality, so it is expected that some LiDAR preprocessing yielding richer geometry information about objects will be required for performant AV decision-making. 

In a real-world driving scenario, AVs will need to cope with unpredictable traffic and changing weather conditions, so trajectory planning should in addition be sensitive to uncertainty in the driving environment.  For this initial work we have chosen to keep the complexity of the model low to simplify the problem; one benefit is that the restricted modelling palette forces basic assumptions to be questioned in pursuit of a solution, as opposed to relying on the modelling apparatus alone.  In our case, we have attempted to stretch assumptions about relevant discretizations, moving away from typical fixed structures.  In future work, however, a probabilistic model checker such as PRISM \cite{kwiatkowska_prism_2011} could be stripped down and implemented on an autonomous agent to handle uncertainty in the local environment. 

The main limitation of our approach is that it is not goal-directed.  While our method is capable of local  tactical planning for obstacle avoidance, any realistic scenario involving AVs (e.g., overtaking) requires the ability to make progress towards a goal relevant to the task.  For example, in our previous work \cite{pagojus_simulation_2021} a simulated LiDAR array provided sensory input sufficient for determining whether an overtake was possible.  Subsequent research will seek to improve on our solution by adding goal-directed behaviour to the autonomous agent as a next step towards transparent and ecologically valid trajectory planning.

The major benefit of our approach is that it is tailored for individual scenarios and takes place in real-time using no pre-computed data \cite{pagojus_simulation_2021}. Furthermore, it makes hard decisions which are in principle transparent.  Popular machine learning approaches for trajectory planning, such as deep reinforcement learning \cite{Josef2020}, are either trained offline on datasets irrelevant to the immediate context, or trained online within a simulated environment.  While predictions are fast in offline methods, they can result in unexpected or risky behaviour for unseen cases, which in autonomous driving can be dangerous, as evidenced in recent high profile accidents like the Tesla crash while in autopilot mode \cite{lee_us_2016} and the Uber autonomous taxi crash \cite{conger2020}, both of which were fatal. Online training of reinforcement learning in a simulated environment might be able to generate a larger variability of situations for training (and a much larger dataset than any real world data), but it would ultimately generate a rigid black box system which would not be transparent or guaranteed to react safely in all situations. Our solution relies on data from the local environment and generates explainable trajectories in real-time.  Unlike machine learning methods, our basic approach makes clear decisions which are transparent by design.

Safety is ensured by our abstraction through over-approximation of the robot workspace and strict adherence to the robot safe zone.  Similar concepts restricting the local behaviour of robots have been used elsewhere.  In \cite{lapierre_guaranteed_2012}, for example, the notion of a \textit{safe maneuvering zone} (SMZ) was used with a kinematic model for obstacle avoidance.  The SMZ defines a circular boundary around the closest detected obstacle, creating a temporary sub-goal which minimally deforms the original robot path.  The contour of any encountered obstacle is navigated until the robot can return to its original path.  However, as the SMZ places safety bounds on obstacles, efficient strategies for obstacle avoidance may be ignored.  Our safe zone is egocentric, so paths are not constrained by the contours of obstacles in the environment.  

Previous work \cite{bergman_improved_2021} combining lattice-based planning with optimal control has informed the development of our method.  Here optimal path planning algorithms were used to generate motion primitives which can then be chained, producing locally optimal solutions to the path planning problem. However, it is a non-trivial task for a robot to precisely determine its position and follow a trajectory, often the approach favoured by safe navigation methods in the literature, such as control barrier functions \cite{singletary_comparative_2020}.  Instead we focus attention on possible collisions which are tracked until out of reach whilst respecting the robot safe zone.  Hence the trajectories produced by sequences of motion primitives (i.e., control tasks) are flexible in our method, as long as the control goal of eliminating disturbances is achieved.

\subsection{Comparison with physics modelling}
In a similar vein to model checking, we have approached closed-loop navigation from a lower level of abstraction; in brief, this consists of representing the robot, its behaviour and the external environment in the physics engine Box2D \cite{CattoErincatto/box2d:Games} (unpublished data). In this framework, we replace the model checking step by simulation of the execution of a task (or sequence of tasks) in the physics engine. The outcome of simulation can then be used to determine a sequence that best satisfies a goal. Modelling through a widely utilised and validated physics library allows for accurate simulation of complex, dynamic environments. On the other hand, the speed of this method is strongly correlated 
with the number of bodies used in the simulation. In simple scenarios, optimal path selection is possible within the LiDAR sampling rate. 
However, cluttered environments and/or the creation of large task trees result in highly variable performance 
to a point where it no longer suits real-world applications in robots with low CPU clock speed (our data was collected on a 1.4GHz CPU). This limitation is not in the model checking approach, making it more robust, reliable and suited to a wider range of scenarios.

\section*{Acknowledgements}
This work was supported by a grant from the UKRI Strategic Priorities Fund to the UKRI Research Node on Trustworthy Autonomous Systems Governance and Regulation [EP/V026607/1, 2020-2024]; the UKRI Centre for Doctoral Training in Socially Intelligent Artificial Agents [EP/S02266X/1]; and the UKRI Engineering and Physical Sciences Research Council Doctoral Training Partnership award [EP/T517896/1-312561-05].

\newpage
\bibliographystyle{eptcs}
\bibliography{references}

\begin{thebibliography}{10}
\providecommand{\bibitemdeclare}[2]{}
\providecommand{\surnamestart}{}
\providecommand{\surnameend}{}
\providecommand{\urlprefix}{Available at }
\providecommand{\url}[1]{\texttt{#1}}
\providecommand{\href}[2]{\texttt{#2}}
\providecommand{\urlalt}[2]{\href{#1}{#2}}
\providecommand{\doi}[1]{doi:\urlalt{https://doi.org/#1}{#1}}
\providecommand{\eprint}[1]{arXiv:\urlalt{https://arxiv.org/abs/#1}{#1}}
\providecommand{\bibinfo}[2]{#2}

\bibitemdeclare{book}{baier_principles_2008}
\bibitem{baier_principles_2008}
\bibinfo{author}{Christel \surnamestart Baier\surnameend} \&
  \bibinfo{author}{Joost-Pieter \surnamestart Katoen\surnameend}
  (\bibinfo{year}{2008}): \emph{\bibinfo{title}{Principles {Of} {Model}
  {Checking}}}.
\newblock \bibinfo{volume}{950}, \bibinfo{publisher}{The MIT Press},
  \bibinfo{address}{Cambridge, Mass}, \doi{10.1093/comjnl/bxp025}.

\newblock \bibinfo{note}{Publication Title: MIT Press ISSN: 00155713}.

\bibitemdeclare{inproceedings}{uppaal-application3}
\bibitem{uppaal-application3}
\bibinfo{author}{Davide \surnamestart Basile\surnameend},
  \bibinfo{author}{Alessandro \surnamestart Fantechi\surnameend} \&
  \bibinfo{author}{Irene \surnamestart Rosadi\surnameend}
  (\bibinfo{year}{2021}): \emph{\bibinfo{title}{Formal Analysis of the UNISIG
  Safety Application Intermediate Sub-layer}}.
\newblock In \bibinfo{editor}{Alberto \surnamestart Lluch~Lafuente\surnameend}
  \& \bibinfo{editor}{Anastasia \surnamestart Mavridou\surnameend}, editors:
  {\slshape \bibinfo{booktitle}{Formal Methods for Industrial Critical
  Systems}}, \bibinfo{publisher}{Springer International Publishing},
  \bibinfo{address}{Cham}, pp. \bibinfo{pages}{174--190},
  \doi{10.1007/978-3-030-85248-1\_11}.

\bibitemdeclare{article}{bergman_improved_2021}
\bibitem{bergman_improved_2021}
\bibinfo{author}{Kristoffer \surnamestart Bergman\surnameend},
  \bibinfo{author}{Oskar \surnamestart Ljungqvist\surnameend} \&
  \bibinfo{author}{Daniel \surnamestart Axehill\surnameend}
  (\bibinfo{year}{2021}): \emph{\bibinfo{title}{Improved {Path} {Planning} by
  {Tightly} {Combining} {Lattice}-{Based} {Path} {Planning} and {Optimal}
  {Control}}}.
\newblock {\slshape \bibinfo{journal}{IEEE Transactions on Intelligent
  Vehicles}} \bibinfo{volume}{6}(\bibinfo{number}{1}), pp.
  \bibinfo{pages}{57--66}, \doi{10.1109/TIV.2020.2991951}.

\bibitemdeclare{book}{Braitenberg1986Vehicles:Psychology}
\bibitem{Braitenberg1986Vehicles:Psychology}
\bibinfo{author}{V.~\surnamestart Braitenberg\surnameend}
  (\bibinfo{year}{1986}): \emph{\bibinfo{title}{{Vehicles: Experiments in
  Synthetic Psychology}}}.
\newblock \bibinfo{publisher}{MIT Press}, \bibinfo{address}{Cambridge,
  Massachussets}.
\newblock
  \urlprefix\url{https://books.google.co.uk/books/about/Vehicles.html?id=7KkUAT_q_sQC&redir_esc=y}.

\bibitemdeclare{article}{cardoso2021}
\bibitem{cardoso2021}
\bibinfo{author}{R.~C. \surnamestart Cardoso\surnameend},
  \bibinfo{author}{G.~\surnamestart Kourtis\surnameend}, \bibinfo{author}{L.~A.
  \surnamestart Dennis\surnameend}, \bibinfo{author}{C.~\surnamestart
  Dixon\surnameend}, \bibinfo{author}{M.~\surnamestart Farrell\surnameend},
  \bibinfo{author}{M.~\surnamestart Fisher\surnameend} \&
  \bibinfo{author}{M.~\surnamestart Webster\surnameend} (\bibinfo{year}{2021}):
  \emph{\bibinfo{title}{A Review of Verification and Validation for Space
  Autonomous Systems}}.
\newblock {\slshape \bibinfo{journal}{Current Robotics Reports}}
  \bibinfo{volume}{2}(\bibinfo{number}{3}), pp. \bibinfo{pages}{273–--283},
  \doi{10.1007/s43154-021-00058-1}.

\bibitemdeclare{misc}{CattoErincatto/box2d:Games}
\bibitem{CattoErincatto/box2d:Games}
\bibinfo{author}{E.~\surnamestart Catto\surnameend}:
  \emph{\bibinfo{title}{{erincatto/box2d: Box2D is a 2D physics engine for
  games}}}.
\newblock \urlprefix\url{https://github.com/erincatto/box2d}.

\bibitemdeclare{article}{conger2020}
\bibitem{conger2020}
\bibinfo{author}{K.~\surnamestart Conger\surnameend} (\bibinfo{year}{2020}):
  \emph{\bibinfo{title}{Driver Charged in Uber’s Fatal 2018 Autonomous Car
  Crash}}.
\newblock {\slshape \bibinfo{journal}{The New York Times}}.
\urlprefix\url{https://www.nytimes.com/2020/09/15/technology/uber-autonomous-crash-driver-charged.html}

\bibitemdeclare{article}{dixit_trajectory_2018}
\bibitem{dixit_trajectory_2018}
\bibinfo{author}{Shilp \surnamestart Dixit\surnameend}, \bibinfo{author}{Saber
  \surnamestart Fallah\surnameend}, \bibinfo{author}{Umberto \surnamestart
  Montanaro\surnameend}, \bibinfo{author}{Mehrdad \surnamestart
  Dianati\surnameend}, \bibinfo{author}{Alan \surnamestart Stevens\surnameend},
  \bibinfo{author}{Francis \surnamestart Mccullough\surnameend} \&
  \bibinfo{author}{Alexandros \surnamestart Mouzakitis\surnameend}
  (\bibinfo{year}{2018}): \emph{\bibinfo{title}{Trajectory planning and
  tracking for autonomous overtaking: {State}-of-the-art and future
  prospects}}.
\newblock {\slshape \bibinfo{journal}{Annual Reviews in Control}}
  \bibinfo{volume}{45}, pp. \bibinfo{pages}{76--86},
  \doi{10.1016/j.arcontrol.2018.02.001}.

\bibitemdeclare{proceedings}{FMAS2021}
\bibitem{FMAS2021}
\bibinfo{editor}{Marie \surnamestart Farrell\surnameend} \&
  \bibinfo{editor}{Matt \surnamestart Luckcuck\surnameend}, editors
  (\bibinfo{year}{2021}): \emph{\bibinfo{title}{Proceedings of the Third
  Workshop on Formal Methods for Autonomous Systems}}. \bibinfo{volume}{348},
  \bibinfo{publisher}{Open Publishing Association}, \doi{10.4204/eptcs.348}.


\bibitemdeclare{inproceedings}{ferrando2018}
\bibitem{ferrando2018}
\bibinfo{author}{Angelo \surnamestart Ferrando\surnameend},
  \bibinfo{author}{Louise~A. \surnamestart Dennis\surnameend},
  \bibinfo{author}{Davide \surnamestart Ancona\surnameend},
  \bibinfo{author}{Michael \surnamestart Fisher\surnameend} \&
  \bibinfo{author}{Viviana \surnamestart Mascardi\surnameend}
  (\bibinfo{year}{2018}): \emph{\bibinfo{title}{Verifying and Validating
  Autonomous Systems: Towards an Integrated Approach}}.
\newblock In \bibinfo{editor}{Christian \surnamestart Colombo\surnameend} \&
  \bibinfo{editor}{Martin \surnamestart Leucker\surnameend}, editors: {\slshape
  \bibinfo{booktitle}{Runtime Verification}}, \bibinfo{publisher}{Springer
  International Publishing}, \bibinfo{address}{Cham}, pp.
  \bibinfo{pages}{263--281}.
  \doi{10.1007/978-3-030-03769-7\_15}

\bibitemdeclare{article}{frgihoirmino2020}
\bibitem{frgihoirmino2020}
\bibinfo{author}{D.~\surnamestart Fraser\surnameend},
  \bibinfo{author}{R.~\surnamestart Giaquinta\surnameend},
  \bibinfo{author}{\surnamestart Hoffmann\surnameend},
  \bibinfo{author}{M.~\surnamestart Ireland\surnameend},
  \bibinfo{author}{A.~\surnamestart Miller\surnameend} \&
  \bibinfo{author}{G.~\surnamestart Norman\surnameend} (\bibinfo{year}{2020}):
  \emph{\bibinfo{title}{Collaborative models for autonomous systems controller
  synthesis}}.
\newblock {\slshape \bibinfo{journal}{Form Aspects of Computing}}
  \bibinfo{volume}{32}, pp. \bibinfo{pages}{157–--186},
  \doi{10.1109/TCST.2006.872519}.

\bibitemdeclare{inproceedings}{hamilton2022}
\bibitem{hamilton2022}
\bibinfo{author}{J.~\surnamestart Hamilton\surnameend},
  \bibinfo{author}{I.~\surnamestart Stefanakos\surnameend},
  \bibinfo{author}{R.~\surnamestart Calinescu\surnameend} \&
  \bibinfo{author}{J.~\surnamestart Cámara\surnameend} (\bibinfo{year}{2022}):
  \emph{\bibinfo{title}{Towards Adaptive Planning of Assistive-care Robot
  Tasks}}.
\newblock In \bibinfo{editor}{Luckcuck} \& \bibinfo{editor}{Farrell}
  \cite{FMAS2022}, pp. \bibinfo{pages}{175--183}, \doi{10.4204/eptcs.371}.

\bibitemdeclare{article}{Havelund2001-3}
\bibitem{Havelund2001-3}
\bibinfo{author}{K.~\surnamestart Havelund\surnameend},
  \bibinfo{author}{M.~\surnamestart Lowry\surnameend} \&
  \bibinfo{author}{J.~\surnamestart Penix\surnameend} (\bibinfo{year}{2001}):
  \emph{\bibinfo{title}{Formal Analysis of a Space-Craft Controller Using
  SPIN.}}
\newblock {\slshape \bibinfo{journal}{Software Engineering, IEEE Transactions
  on}} \bibinfo{volume}{27}, pp. \bibinfo{pages}{749--765},
  \doi{10.1109/32.940728}.

\bibitemdeclare{book}{holz2011}
\bibitem{holz2011}
\bibinfo{author}{G.~\surnamestart Holzmann\surnameend} (\bibinfo{year}{2011}):
  \emph{\bibinfo{title}{The {SPIN} Model Checker: Primer and Reference
  Manual}}, \bibinfo{edition}{1st} edition.
\newblock \bibinfo{publisher}{Addison-Wesley Professional}.

\bibitemdeclare{article}{Josef2020}
\bibitem{Josef2020}
\bibinfo{author}{S.~\surnamestart Josef\surnameend} \&
  \bibinfo{author}{A.~\surnamestart Degani\surnameend} (\bibinfo{year}{2020}):
  \emph{\bibinfo{title}{Deep Reinforcement Learning for Safe Local Planning of
  a Ground Vehicle in Unknown Rough Terrain}}.
\newblock {\slshape \bibinfo{journal}{IEEE Robotics and Automation Letters}}
  \bibinfo{volume}{5}(\bibinfo{number}{4}), pp. \bibinfo{pages}{6748--6755},
  \doi{10.1109/LRA.2020.3011912}.

\bibitemdeclare{inproceedings}{kwiatkowska_prism_2011}
\bibitem{kwiatkowska_prism_2011}
\bibinfo{author}{M.~\surnamestart Kwiatkowska\surnameend},
  \bibinfo{author}{G.~\surnamestart Norman\surnameend} \&
  \bibinfo{author}{D.~\surnamestart Parker\surnameend} (\bibinfo{year}{2011}):
  \emph{\bibinfo{title}{{PRISM} 4.0: {Verification} of {Probabilistic}
  {Real}-{Time} {Systems}}}.
\newblock In \bibinfo{editor}{Ganesh \surnamestart Gopalakrishnan\surnameend}
  \& \bibinfo{editor}{Shaz \surnamestart Qadeer\surnameend}, editors: {\slshape
  \bibinfo{booktitle}{Computer Aided Verification}},
  \bibinfo{publisher}{Springer Berlin Heidelberg}, \bibinfo{address}{Berlin,
  Heidelberg}, pp. \bibinfo{pages}{585--591},
  \doi{10.1007/978-3-642-22110-1\_47}.

\bibitemdeclare{article}{kwiatkowska2022}
\bibitem{kwiatkowska2022}
\bibinfo{author}{M.~\surnamestart Kwiatkowska\surnameend},
  \bibinfo{author}{G.~\surnamestart Norman\surnameend} \&
  \bibinfo{author}{D.~\surnamestart Parker\surnameend} (\bibinfo{year}{2022}):
  \emph{\bibinfo{title}{Probabilistic Model checking and autonomy}}.
\newblock {\slshape \bibinfo{journal}{Annual review of control, robotics, and
  autonomous systems}} \bibinfo{volume}{5}(\bibinfo{number}{1}), pp.
  \bibinfo{pages}{385--410}.
  \doi{10.1146/annurev-control-042820-010947}

\bibitemdeclare{article}{lapierre_guaranteed_2012}
\bibitem{lapierre_guaranteed_2012}
\bibinfo{author}{Lionel \surnamestart Lapierre\surnameend} \&
  \bibinfo{author}{Rene \surnamestart Zapata\surnameend}
  (\bibinfo{year}{2012}): \emph{\bibinfo{title}{A guaranteed obstacle avoidance
  guidance system: {The} safe maneuvering zone}}.
\newblock {\slshape \bibinfo{journal}{Autonomous Robots}}
  \bibinfo{volume}{32}(\bibinfo{number}{3}), pp. \bibinfo{pages}{177--187},
  \doi{10.1007/s10514-011-9269-5}.

\bibitemdeclare{misc}{LeCun2022AOpenReview}
\bibitem{LeCun2022AOpenReview}
\bibinfo{author}{Yann \surnamestart LeCun\surnameend} (\bibinfo{year}{2022}):
  \emph{\bibinfo{title}{{A Path Towards Autonomous Machine Intelligence |
  OpenReview}}}.
\newblock \urlprefix\url{https://openreview.net/forum?id=BZ5a1r-kVsf}.

\bibitemdeclare{misc}{lee_us_2016}
\bibitem{lee_us_2016}
\bibinfo{author}{David \surnamestart Lee\surnameend} (\bibinfo{year}{2016}):
  \emph{\bibinfo{title}{{US} opens investigation into {Tesla} after fatal
  crash}}.
\newblock \urlprefix\url{https://www.bbc.co.uk/news/technology-36680043}.

\bibitemdeclare{inproceedings}{lehmannneedles2021}
\bibitem{lehmannneedles2021}
\bibinfo{author}{S.~\surnamestart Lehmann\surnameend},
  \bibinfo{author}{A.~\surnamestart Rogalla\surnameend},
  \bibinfo{author}{M.~\surnamestart Neidhardt\surnameend},
  \bibinfo{author}{A.~\surnamestart Schlaefer\surnameend} \&
  \bibinfo{author}{S.~\surnamestart Schupp\surnameend} (\bibinfo{year}{2021}):
  \emph{\bibinfo{title}{Online Strategy Synthesis for Safe and Optimized
  Control of Steerable Needles}}.
\newblock In \bibinfo{editor}{Farrell} \& \bibinfo{editor}{Luckcuck}
  \cite{FMAS2021}, pp. \bibinfo{pages}{128--135}, \doi{10.4204/EPTCS.348.9}.

\bibitemdeclare{article}{li2016}
\bibitem{li2016}
\bibinfo{author}{X.~\surnamestart Li\surnameend},
  \bibinfo{author}{Z.~\surnamestart Sun\surnameend},
  \bibinfo{author}{D.~\surnamestart Cao\surnameend},
  \bibinfo{author}{Z.~\surnamestart He\surnameend} \&
  \bibinfo{author}{Q.~\surnamestart Zhu\surnameend} (\bibinfo{year}{2016}):
  \emph{\bibinfo{title}{Real-time trajectory planning for autonomous urban
  driving: Framework, algorithms, and verifications}}.
\newblock {\slshape \bibinfo{journal}{{IEEE/ASME} Transactions on
  Mechatronics}} \bibinfo{volume}{21}(\bibinfo{number}{2}), pp.
  \bibinfo{pages}{740--753}, \doi{10.1109/TMECH.2015.2493980}.

\bibitemdeclare{article}{Dennis2016}
\bibitem{Dennis2016}
\bibinfo{author}{Dennis \surnamestart Louise\surnameend},
  \bibinfo{author}{Michael \surnamestart Fisher\surnameend},
  \bibinfo{author}{Nicholas \surnamestart Lincoln\surnameend},
  \bibinfo{author}{Alexei \surnamestart Lisitsa\surnameend} \&
  \bibinfo{author}{Sandor \surnamestart Veres\surnameend}
  (\bibinfo{year}{2016}): \emph{\bibinfo{title}{Practical verification of
  decision-making in agent-based autonomous systems}}.
\newblock {\slshape \bibinfo{journal}{Automated Software Engineering}}
  \bibinfo{volume}{23}(\bibinfo{number}{3}), pp. \bibinfo{pages}{305--359},
  \doi{10.1007/s10515-014-0168-9}.

\bibitemdeclare{inproceedings}{lu2015}
\bibitem{lu2015}
\bibinfo{author}{Y.~\surnamestart Lu\surnameend},
  \bibinfo{author}{A.~\surnamestart Miller\surnameend},
  \bibinfo{author}{C.~\surnamestart Johnson\surnameend},
  \bibinfo{author}{Z.~\surnamestart Peng\surnameend} \&
  \bibinfo{author}{T.~\surnamestart Zhao\surnameend} (\bibinfo{year}{2014}):
  \emph{\bibinfo{title}{Availability analysis of satellite positioning systems
  for avaiation using the {P}rism model checker}}.
\newblock In: {\slshape \bibinfo{booktitle}{Proceedings of the 17th
  International Conference on Computational Science and Engineering ({CSE}
  2014)}}, pp. \bibinfo{pages}{704--713}, \doi{10.1109/CSE.2014.148}.

\bibitemdeclare{proceedings}{FMAS2022}
\bibitem{FMAS2022}
\bibinfo{editor}{Matt \surnamestart Luckcuck\surnameend} \&
  \bibinfo{editor}{Marie \surnamestart Farrell\surnameend}, editors
  (\bibinfo{year}{2022}): \emph{\bibinfo{title}{Proceedings of the Fourth
  International Workshop on Formal Methods for Autonomous Systems (FMAS) and
  Fourth International Workshop on Automated and verifiable Software sYstem
  DEvelopment (ASYDE)}}. \bibinfo{volume}{371}, \bibinfo{publisher}{Open
  Publishing Association}, \doi{10.4204/eptcs.371}.

\bibitemdeclare{inproceedings}{pagojus_simulation_2021}
\bibitem{pagojus_simulation_2021}
\bibinfo{author}{Daumantas \surnamestart Pagojus\surnameend},
  \bibinfo{author}{Alice \surnamestart Miller\surnameend},
  \bibinfo{author}{Bernd \surnamestart Porr\surnameend} \&
  \bibinfo{author}{Ivaylo \surnamestart Valkov\surnameend}
  (\bibinfo{year}{2021}): \emph{\bibinfo{title}{Simulation and {Model}
  {Checking} for {Close} to {Realtime} {Overtaking} {Planning}}}.
\newblock In \bibinfo{editor}{Farrell} \& \bibinfo{editor}{Luckcuck}
  \cite{FMAS2021}, pp. \bibinfo{pages}{20--37}, \doi{10.4204/EPTCS.348.2}.

\bibitemdeclare{book}{siciliano_springer_2016}
\bibitem{siciliano_springer_2016}
\bibinfo{editor}{Bruno \surnamestart Siciliano\surnameend} \&
  \bibinfo{editor}{Oussama \surnamestart Khatib\surnameend}, editors
  (\bibinfo{year}{2016}): \emph{\bibinfo{title}{Springer {Handbook} of
  {Robotics}}}.
\newblock \bibinfo{publisher}{Springer International Publishing},
  \bibinfo{address}{Cham}, \doi{10.1007/978-3-319-32552-1}.

\bibitemdeclare{misc}{singletary_comparative_2020}
\bibitem{singletary_comparative_2020}
\bibinfo{author}{Andrew \surnamestart Singletary\surnameend},
  \bibinfo{author}{Karl \surnamestart Klingebiel\surnameend},
  \bibinfo{author}{Joseph \surnamestart Bourne\surnameend},
  \bibinfo{author}{Andrew \surnamestart Browning\surnameend},
  \bibinfo{author}{Phil \surnamestart Tokumaru\surnameend} \&
  \bibinfo{author}{Aaron \surnamestart Ames\surnameend} (\bibinfo{year}{2020}):
  \emph{\bibinfo{title}{Comparative {Analysis} of {Control} {Barrier}
  {Functions} and {Artificial} {Potential} {Fields} for {Obstacle}
  {Avoidance}}}.
\newblock \urlprefix\url{http://arxiv.org/abs/2010.09819}.
\newblock \bibinfo{note}{ArXiv:2010.09819 [cs, eess]}.

\bibitemdeclare{article}{Spelke2007CoreKnowledge}
\bibitem{Spelke2007CoreKnowledge}
\bibinfo{author}{Elizabeth~S. \surnamestart Spelke\surnameend} \&
  \bibinfo{author}{Katherine~D. \surnamestart Kinzler\surnameend}
  (\bibinfo{year}{2007}): \emph{\bibinfo{title}{{Core knowledge}}}.
\newblock {\slshape \bibinfo{journal}{Developmental Science}}
  \bibinfo{volume}{10}(\bibinfo{number}{1}), pp. \bibinfo{pages}{89--96},
  \doi{10.1111/J.1467-7687.2007.00569.X}.

\bibitemdeclare{inproceedings}{android}
\bibitem{android}
\bibinfo{author}{Güliz \surnamestart Tuncay\surnameend},
  \bibinfo{author}{Soteris \surnamestart Demetriou\surnameend},
  \bibinfo{author}{Karan \surnamestart Ganju\surnameend} \&
  \bibinfo{author}{Carl \surnamestart A.~Gunter\surnameend}
  (\bibinfo{year}{2018}): \emph{\bibinfo{title}{Resolving the Predicament of
  Android Custom Permissions}}.
\newblock In: {\slshape \bibinfo{booktitle}{Network and Distributed System
  Security Symposium}}, pp. \bibinfo{pages}{1--15},
  \doi{10.14722/ndss.2018.23221}.

\bibitemdeclare{article}{Wang2018}
\bibitem{Wang2018}
\bibinfo{author}{L.~\surnamestart Wang\surnameend} \&
  \bibinfo{author}{F.~\surnamestart Cai\surnameend} (\bibinfo{year}{2017}):
  \emph{\bibinfo{title}{{Reliability analysis for flight control systems using
  probabilistic model checking}}}.
\newblock {\slshape \bibinfo{journal}{Proceedings of the {IEEE} International
  Conference on Software Engineering and Service Sciences, {ICSESS}}}
  \bibinfo{volume}{2017-Novem}, pp. \bibinfo{pages}{161--164},
  \doi{10.1109/RAM.2017.7889773}.

\bibitemdeclare{inproceedings}{weissmann2011}
\bibitem{weissmann2011}
\bibinfo{author}{M.~\surnamestart Weißmann\surnameend},
  \bibinfo{author}{S.~\surnamestart Bedenk\surnameend},
  \bibinfo{author}{C.~\surnamestart Buckl\surnameend} \&
  \bibinfo{author}{A.~\surnamestart Knoll\surnameend} (\bibinfo{year}{2011}):
  \emph{\bibinfo{title}{Model Checking Industrial Robot Systems}}.
\newblock In: {\slshape \bibinfo{booktitle}{Model checking software ({SPIN}
  2011)}}, \bibinfo{volume}{6823}, \bibinfo{publisher}{Springer Berlin
  Heidelberg}, pp. \bibinfo{pages}{161--176},
  \doi{10.1007/978-3-642-22306-8\_11}.

\bibitemdeclare{inproceedings}{Yang_2022_ros}
\bibitem{Yang_2022_ros}
\bibinfo{author}{Yi~\surnamestart Yang\surnameend} \& \bibinfo{author}{Tom
  \surnamestart Holvoet\surnameend} (\bibinfo{year}{2022}):
  \emph{\bibinfo{title}{Generating Safe Autonomous Decision-Making in ROS}}.
\newblock In \bibinfo{editor}{Luckcuck} \& \bibinfo{editor}{Farrell} \cite{FMAS2022}, pp. \bibinfo{pages}{184--192}, \doi{10.4204/eptcs.371.13}.


\end{thebibliography}


\end{document}